%% file: main.tex
\documentclass[conference]{IEEEtran}

\usepackage{cite}

\usepackage[T1]{fontenc}
\usepackage[utf8]{inputenc}

\usepackage{graphicx}
\graphicspath{{figures/}}
\usepackage{booktabs}
\usepackage{array}
\usepackage{multirow}

\usepackage{amsmath,amssymb}
\usepackage{listings}
\usepackage{xcolor}
\usepackage[hidelinks]{hyperref}
\usepackage{cleveref}

\usepackage{xspace}

\newcommand{\toolname}{\textsc{Comprendia}\xspace}
\newcommand{\gacp}{\textsc{GACP}\xspace}

\usepackage{balance}

\begin{document}
\bstctlcite{IEEEexample:BSTcontrol}

\title{\toolname: \\AI-Augmented Code Comprehension}

\author{
\IEEEauthorblockN{Costain Nachuma}
\IEEEauthorblockA{Department of Computer Science\\
                  Idaho State University\\
                  Pocatello, Idaho, USA\\
                  costainnachuma@isu.edu}
\and
\IEEEauthorblockN{Minhaz F. Zibran}
\IEEEauthorblockA{Department of Computer Science\\
                  Idaho State University\\
                  Pocatello, Idaho, USA\\
                  zibran@isu.edu}
}

\maketitle

\begin{abstract}
\toolname{} is an Eclipse plugin that integrates structural dependency
visualization with LLM-powered code explanation on a shared interactive
graph for Java program comprehension. The tool rests on four pillars:
(1)~a multi-edge-type dependency graph with live search and multiple
layouts; (2)~LLM explanations grounded in Graph-Aware Callee Pruning
(\gacp), an auditable strategy that selects relevant
callees using the same graph the developer navigates; (3)~a
clone-detection overlay that highlights duplication and suggests
extract-to-parent refactoring opportunities; and (4)~a CVE risk overlay
powered by OSV.dev. \gacp{} uses graph distance, inheritance collapse,
and edge-type weighting to produce prompts that are reproducible across
LLM families and traceable to visible graph nodes. We demonstrate
\toolname{} on a Java project containing known clones and
vulnerabilities, showing how the unified graph substrate supports
comprehension while keeping the developer in
control. Screencast: \url{https://youtu.be/1wlh_RYehzA}
\end{abstract}

\vspace{0.2cm}

\input{sections/01-intro}
\input{sections/02-related}
\input{sections/03-tool}
\input{sections/04-algorithm}
\input{sections/05-evaluation}
\input{sections/06-limitations}

\input{sections/07-conclusion}

\balance
\bibliographystyle{IEEEtran}
\bibliography{refs}

\end{document}

%% file: sections/01-intro.tex
\section{Introduction}
\label{sec:intro}

Modern developers rely increasingly on large language models (LLMs) to
write code~\cite{nachuma2025analyzing, li2025riseaiteammatessoftware}. 
Likewise, they are using LLMs to explain unfamiliar code regions inside their integrated development environment (IDE). The dominant interaction paradigm, shared by Copilot
Chat, Cursor, and Claude Code, treats the developer's text selection as
the LLM's entire view of the
project~\cite{copilot2024,cursor2024}. This is empirically
insufficient: even short methods routinely call into project-internal
helpers whose behaviour is essential to a complete explanation, and a
selection-only prompt forces the model to guess at those helpers from
their names alone. Recent empirical work on AI-assisted programming
finds that the gap between what the model is shown and what it needs to
reason over is a frequent source of hallucinated identifiers and shallow,
name-driven explanations~\cite{champa2024chatgpt,
nachuma2026aiteammates}, in line with longstanding findings on the
cognitive demands of code comprehension~\cite{siegmund2014understanding,latoza2010hard,scalabrino2018readability}.

Tools that attempt to close this gap rely on embedding-similarity
retrieval~\cite{cody2024,liang2024repofuse,tufano2022neuralreview} or
perplexity-based prompt
compression~\cite{jiang2023llmlingua,xu2024recomp,jiang2024longllmlingua,terragni2025llmloop}.
Both families share three limitations: their selection criteria are
\emph{approximate} (a similarity score, not a structural guarantee),
\emph{not reproducible} across LLM families, and \emph{not auditable}
by the developer. We argue that the IDE's own dependency graph, which
the developer can see, navigate, and trust, should be the relevance
signal.

This paper presents \toolname, an Eclipse plugin for software
comprehension that integrates dependency-graph visualization,
LLM-assisted code explanation, and clone-refactoring discovery onto a
single graph substrate. \toolname{} makes three contributions, each
grounded in the team's prior published empirical
work~\cite{champa2024chatgpt,nachuma2025maven,nachuma2026aiteammates}
and unified by a single design principle: \emph{the visualization the
developer sees is the data structure the tool reasons over.}

\paragraph{Contribution~1: A multi-edge-type structural substrate}
\toolname{} renders an Eclipse Java project's dependency graph as an
interactive visualization with five structural edge types
(inheritance, interface, field-type, method-parameter, method-return).
Edge-type filters, four layout algorithms, and
in-graph search are exposed through a single toolbar. This extends
Zibran's management-oriented~\cite{Zibran-CloMan-2012, Roy-Keynote-2014, Zibran-CloneFramework-2016} clone-visualization line~\cite{zibran2015cloman} from a
single-edge overlay to a comprehensive structural graph, complementing
code-city work~\cite{wettel2008codecity} and industrial
program-comprehension research~\cite{quante2008comprehension}.

\paragraph{Contribution~2: Graph-Aware Callee Pruning (\gacp)}
\toolname{} introduces \gacp, an algorithm that uses the
same dependency graph as the relevance signal for prompt construction.
\gacp{} (i)~admits callees within $k$~graph hops, (ii)~collapses
near-duplicate siblings sharing an inheritance ancestor into a single
delegate, and (iii)~allocates the remaining budget under an
edge-type-weighted score. Every emitted callee carries an
\emph{inclusion reason} mapping one-to-one onto a visible graph node,
making decisions \emph{auditable} and
\emph{reproducible}. \Cref{sec:algorithm} gives the full specification.

\paragraph{Contribution~3: Clone-refactoring discovery as a graph overlay}
\toolname{} integrates the \textsc{CloMan} clone-detection
engine~\cite{zibran2015cloman} with the dependency graph by painting
clone-group members as red-highlighted nodes. Clicking a group focuses
the graph on its members and reveals extract-to-parent refactoring
opportunities~\cite{alomar2022refactoring}. Because the overlay shares
the same substrate, clicking a cloned class yields both its graph
context and a \gacp-enriched explanation via the inheritance-collapse
phase.

All three contributions share a single substrate: the project dependency
graph that \gacp{} queries, the clone overlay paints onto, and the
developer reads to verify the tool's reasoning. \toolname{} also
exposes a CVE risk badge via
OSV.dev~\cite{nachuma2025maven,improta2024llmsecurity}, but we exclude
this from the contribution claim as its graph integration is forthcoming
work. The plugin, benchmark, and replay harness are publicly
available~\cite{comprendiaArtifact}.

%% file: sections/02-related.tex
\section{Related Work}
\label{sec:related}

\subsection{IDE-Native LLM Code Explanation}

Several tools embed LLM-powered code explanation directly in the
developer editor. GitHub Copilot
Chat~\cite{copilot2024} and Cursor~\cite{cursor2024} send the text selection for the developer to the model with no additional project
context; the model must infer callee behaviour from names alone. Sourcegraph Cody~\cite{cody2024} and   RepoFuse~\cite{liang2024repofuse} go further by retrieving   project-internal code via embedding similarity, ranking candidate snippets by cosine distance to the selection. In all cases the inclusion decision is approximate (a similarity score above a threshold), not reproducible across embedding models, and not auditable by the developer: there is no visible structure the user can inspect to verify why a snippet was included. Patel et al.~\cite{patel} offer a valuable conceptual framework, arguing that unreviewed AI-generated code causes long-term \emph{knowledge erosion}; \gacp's auditable citations operationalize a concrete,
tool-level countermeasure to exactly this risk.

\subsection{Prompt Compression for Long Contexts}

A complementary line of work starts from too much context and   compresses it. LLMLingua~\cite{jiang2023llmlingua} and its long-context extension LongLLMLingua~\cite{jiang2024longllmlingua} use a small language model's perplexity scores to drop low-information tokens, with   question-aware compression for long contexts.   RECOMP~\cite{xu2024recomp} trains extractive and abstractive compressors   to select or rewrite retrieved passages before they enter the prompt. These techniques effectively reduce token costs, but their compression   decisions are model-internal: different perplexity estimators or   compressor checkpoints yield different prompts, and the developer   cannot trace a retained token back to a project-level structure.   \gacp{} differs in kind: its only relevance signal is the dependency   graph the developer already sees, making every inclusion decision   traceable to a visible graph node and reproducible across LLM families.

\subsection{Structural Visualization in IDEs}

Structural visualization tools help developers navigate large   codebases. CodeCity renders software metrics as 3D   cityscapes~\cite{wettel2008codecity}; Lattix, Structure101, and Eclipse   Zest (used by Project Usus~\cite{usus2017}) render dependency matrices,   layered diagrams, or general-purpose graphs within the Eclipse   platform. Quante~\cite{quante2008comprehension} evaluates dynamic   object-process graphs for industrial program comprehension.   Zibran~\cite{zibran2015cloman} proposes clone analysis and   visualization with respect to inheritance hierarchies and call graphs   as necessary support for clone refactoring. \toolname{} continues this   line: it extends Zibran's clone-visualization substrate from a   standalone clone view to a multi-edge-type dependency graph that   supports navigation, AI explanation, and refactoring discovery on the   same surface. None of the tools above combine structural visualization   with LLM-driven code explanation.
\subsection{Position of \toolname{}}

\Cref{tab:position} summarises the gap. \toolname{} is, to our knowledge, among the first IDE-native systems to use the project's dependency graph as the relevance signal for prompt construction,  making callee-inclusion decisions auditable and reproducible.

  \begin{table}[t]
    \centering
    \caption{Position of \toolname{} among related approaches.}
    \label{tab:position}
    \renewcommand{\arraystretch}{1.0}
    \footnotesize
    \begin{tabular}{@{}lc@{}c@{}cc@{}}
      \toprule
      & \rotatebox{45}{AI-explanation}
      & \rotatebox{45}{Graph-grounded}
      & \rotatebox{45}{IDE-native} 
      & \rotatebox{45}{Auditable} \\
      \midrule
      Copilot Chat / Cursor  & \checkmark & --  & \checkmark & -- \\
      Cody / RepoFuse        & \checkmark & --  & \checkmark & -- \\
      LLMLingua / RECOMP     & --         & --  & --         & -- \\
      CodeCity / Zest         & --        & \checkmark & \checkmark & \checkmark \\
      \textbf{\toolname{} (\gacp)} & \checkmark & \checkmark & \checkmark & \checkmark \\
      \bottomrule
    \end{tabular}
  \end{table}

%% file: sections/03-tool.tex
\section{Tool Overview}
\label{sec:tool}

\subsection{User-Facing Workflow}

A developer opens an unfamiliar Eclipse Java project and right-clicks
\emph{Visualize Dependencies}. \toolname{} parses the workspace
incrementally via JDT and renders a dependency graph inside the IDE:
nodes are classes, interfaces, and enums; edges encode five structural
relationships (inheritance, interface, field-type, method-parameter,
method-return). The developer toggles edge types, switches among four
layout algorithms (spring, tree, radial, grid), and uses in-graph search
to locate a class of interest. Double-clicking a node opens the
corresponding source file.

To understand a method, the developer highlights its body and clicks
\emph{Explain with AI}. \toolname{} invokes \gacp{}
(\Cref{sec:algorithm}) to select project-internal callees from the
graph, assembles a prompt containing the selection and the chosen callee
snippets, and routes the request to the configured LLM backend. The
explanation appears in the InsightPanel with callee citations the
developer can click to navigate to the referenced source. Because each
cited callee maps one-to-one onto a visible graph node, the developer
can verify the tool's reasoning against the same structure used to
produce it.

Two overlays enrich the graph without leaving it. Running \emph{Detect
Clones} invokes the CloMan clone-detection
engine~\cite{zibran2015cloman} via a reflection bridge; clone-group
members are highlighted in red, and clicking a group focuses the graph
on its members, exposing candidate extract-to-parent refactoring
opportunities. Independently, parsing the project's \texttt{pom.xml}
and querying the OSV.dev vulnerability database~\cite{osvdev} paints
CVE-severity borders on affected nodes (red for critical, orange for
high), grounded in our prior finding that 62.89\% of latest Maven
releases carry transitive
vulnerabilities~\cite{nachuma2025maven}.

\begin{figure}[t]
  \centering
  \includegraphics[width=\columnwidth]{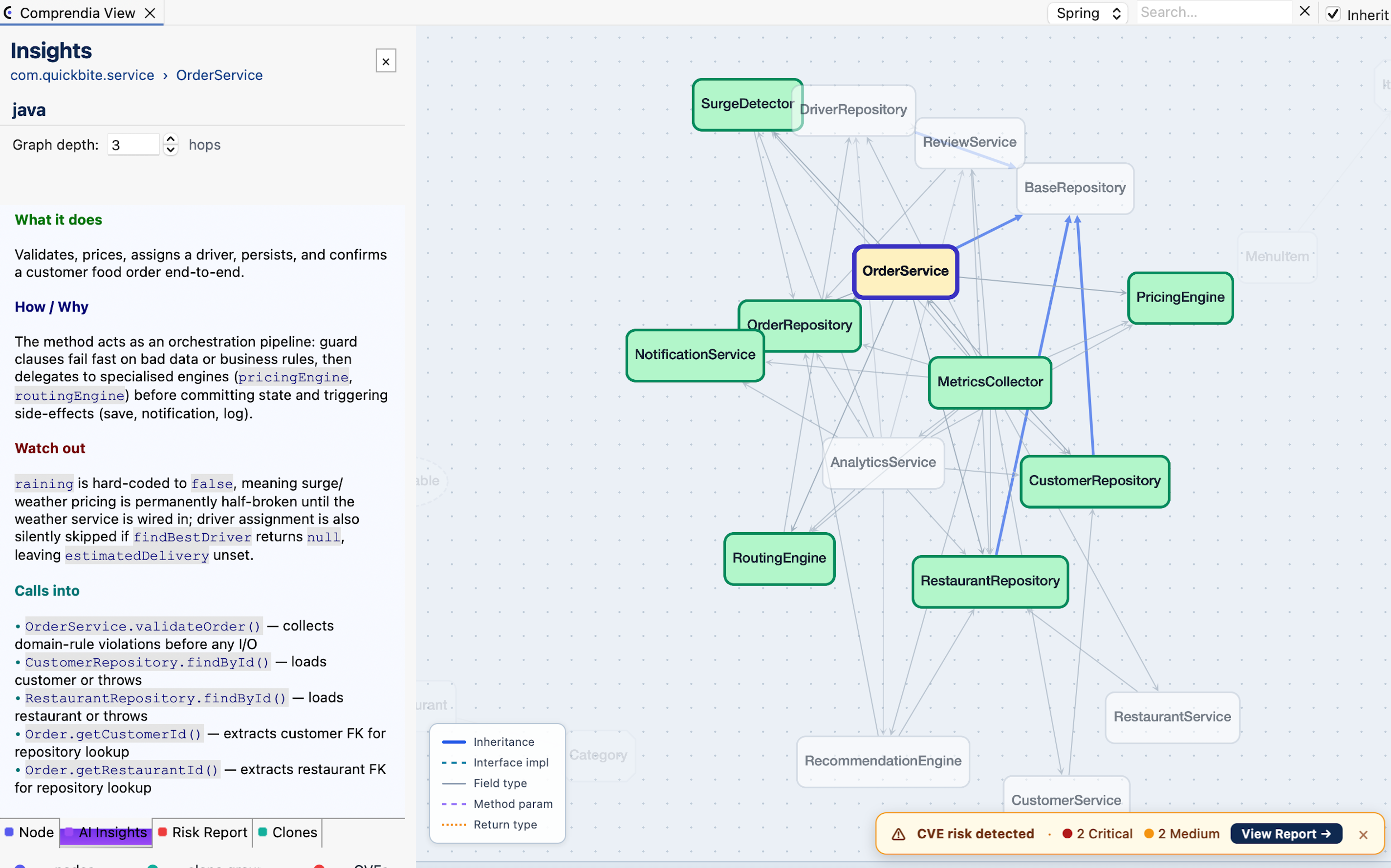}
  \caption{\toolname{} showing the dependency graph with \texttt{OrderService} selected. The InsightPanel (left) displays the GACP-grounded AI explanation with callee citations. The toolbar (top) exposes edge-type filters, layout selection, and search. The CVE risk banner (bottom) reports detected vulnerabilities.}
  \label{fig:ui}
\end{figure}

\begin{figure}[t]
  \centering
  \includegraphics[width=\columnwidth]{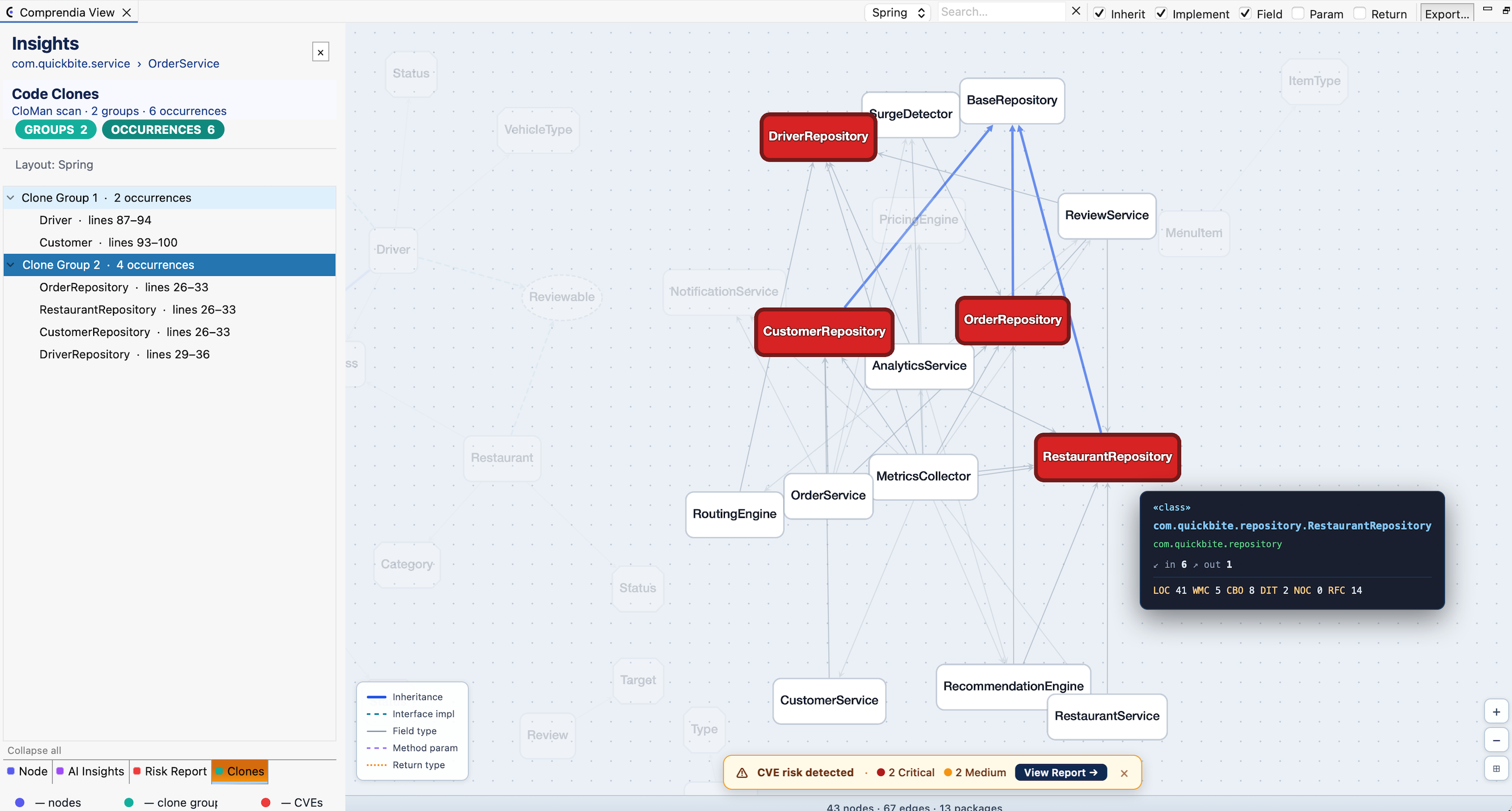}
  \caption{Clone overlay: CloMan detects two clone groups (6~occurrences). The four-class \texttt{BaseRepository} group ($k{=}4$) is highlighted in red; non-clone nodes are grayed. The tooltip shows code metrics for the selected member. This is the group \gacp's Phase~2 collapses in \Cref{sec:evaluation}.}
  \label{fig:clones}
\end{figure}

\subsection{Architecture and Infrastructure}

\toolname{} is organized in five layers. The \emph{UI layer} hosts the
graph view (Cytoscape.js~\cite{franz2016cytoscapejs} in an SWT Browser),
the InsightPanel, and the toolbar. The \emph{analysis layer} builds the
graph via JDT AST parsing~\cite{eclipse_jdt}, runs \gacp, and bridges
CloMan for clone detection. The \emph{AI layer} exposes a single
\texttt{AiGateway} interface with three production backends (Anthropic
Claude, OpenAI, Ollama) plus a deterministic mock for offline use; all
receive the same \gacp-assembled prompt, enabling the cross-LLM
comparison in \Cref{sec:evaluation}. The \emph{data layer} reads JDT
models, queries OSV.dev~\cite{osvdev} for CVE data, and accesses git
history via JGit~\cite{eclipse_jgit}. An opt-in telemetry service logs
feature activations and task timestamps to a local JSONL file for
controlled user studies~\cite{basha2025codewatcher}.

\subsection{Practical Use Scenarios}

\toolname{} targets three recurring needs. A developer onboarding onto
an unfamiliar module clicks \emph{Explain with AI} and receives an
explanation whose callee citations resolve to the graph already on
screen, rather than an opaque summary. A maintainer preparing a
refactoring sprint runs \emph{Detect Clones} to surface duplicate
implementations and their common-ancestor candidates before writing
code. A team triaging a dependency advisory uses the CVE overlay to see
which classes touch a flagged package. For researchers, \gacp's
parameterized, auditable selection offers a controllable variable for
studying how context-selection strategy affects LLM-assisted
comprehension.

%% file: sections/04-algorithm.tex
%
\section{Graph-Aware Callee Pruning}
\label{sec:algorithm}

This section specifies \gacp, a pure, static function that chooses
project-internal callees for the LLM-explanation prompt. Given the
enclosing class FQN, AST-extracted callees (filtered to project-internal
symbols via JDT), a read-only \texttt{GraphIndex} over the dependency
graph, and a token \texttt{Budget}, it returns a ranked list of
\texttt{GacpCallee} records. Each record carries the callee's FQN,
graph distance, snippet mode (\texttt{FULL\_BODY} or
\texttt{SIGNATURE\_ONLY}), rendered prompt fragment, and a structured
\texttt{inclusionReason} (e.g., \emph{``distance-1 INHERITANCE; full
body''}) that maps one-to-one onto a visible graph node. The function
contains no randomness, no LLM calls, and no UI dependencies.

\subsection{Phase~1: Reachability Filter}
\label{sec:algorithm:phase1}

Phase~1 admits a candidate callee if its declaring class lies within
$K_{\max}{=}2$~hops from the enclosing class on the dependency graph;
self-calls (distance~0) are admitted directly. The cutoff is a hard
structural admission rule, not a soft penalty. Two hops covers a
selection's direct callees and their immediate structural context
while bounding prompt size; deeper callees add tokens faster than
comprehension signal. Candidates are
de-duplicated by declaring class so that Phases~2 and~3 operate at
owner granularity. The snippet mode is distance-dependent:
distance-${\leq}1$ callees are rendered as \texttt{FULL\_BODY},
distance-2 as \texttt{SIGNATURE\_ONLY}.

\subsection{Phase~2: Topological Redundancy Elimination}
\label{sec:algorithm:phase2}

Any group of ${\geq}2$ surviving candidates whose declaring classes
share an \emph{immediate} inheritance ancestor~$A$ is replaced with
$A$ itself (rendered as \texttt{FULL\_BODY}) plus a single concrete
delegate. The ``immediate'' qualifier guards against false collapses
through \texttt{java.lang.Object}. This phase yields the largest savings on clone-touching selections.
For example, when the developer requests an explanation inside one of
four sibling repository classes sharing \texttt{BaseRepository},
Phase~2 collapses the four near-duplicate method bodies into the
\texttt{BaseRepository} contract (rendered in full) plus one concrete
delegate, producing an explanation that describes the shared contract
rather than repeating sibling implementations
(\Cref{sec:evaluation}).

\subsection{Phase~3: Edge-Type-Weighted Budget Allocation}
\label{sec:algorithm:phase3}

The final phase ranks survivors by edge-type weight divided by squared
graph distance, favoring inheritance and interface edges (weight~$1.0$)
over field-type ($0.6$) and method-parameter or method-return ($0.4$)
relationships. Candidates are emitted greedily under the token budget,
skipping degenerate snippets. Given fixed inputs, \gacp{} returns the same
output on every invocation: Phase~1 is $O(|V|{+}|E|)$; Phases~2
and~3 are $O(|C| \log |C|)$ on the surviving set, which is
single-digit in practice. \Cref{sec:evaluation} confirms identical
prompts across three LLM families.

%% file: sections/05-evaluation.tex
\section{Demonstration}
\label{sec:evaluation}

We illustrate \gacp's behaviour on six method-body selections from the
\textsc{quickbite} open-source Java benchmark (33~classes, 6~packages),
replayed against three LLM families (Claude Sonnet~4.6, GPT-4o-mini,
Llama~3 via Ollama). The harness compares four modes per fixture:
\textsc{Selection-only} (no callee context),
\textsc{Legacy} (dedup-and-cap at 5 callees),
\textsc{Gacp-sig} (GACP with signature-only rendering), and
\textsc{Gacp-body} (GACP with full bodies for distance-1 callees, the
shipping default). All modes share the same JDT AST visitor and prompt
scaffolding; they differ only in callee filtering and rendering.
\Cref{tab:fixtures} lists the six fixtures: three exercise Phase~1's
reachability filter without engaging collapse; three engage Phase~2 with
group sizes $k = 2, 3, 4$.

\begin{table}[t]
  \centering
  \caption{Benchmark fixtures from \textsc{quickbite}.}
  \label{tab:fixtures}
  \renewcommand{\arraystretch}{1.15}
  \footnotesize
  \begin{tabular}{lll}
    \toprule
    Selection & Category & Calls into \\
    \midrule
    \texttt{placeOrder}            & no-collapse  & 7 helpers across 4 pkgs \\
    \texttt{recommendationEngine}  & no-collapse  & 5 model + util helpers \\
    \texttt{pricingEngineSurge}    & no-collapse  & 4 engine/util helpers \\
    \texttt{baseRepoCollapse}      & collapse $k{=}4$ & 4 sibling repositories \\
    \texttt{reviewableCollapse}    & collapse $k{=}3$ & 3 \texttt{Reviewable}s \\
    \texttt{abstractUserCollapse}  & collapse $k{=}2$ & 2 user subclasses \\
    \bottomrule
  \end{tabular}
\end{table}

\subsection{Token Cost}

\Cref{tab:prompt-tokens} reports prompt-token cost per fixture.
Values are identical across all three LLM families, confirming that the
algorithm's output is gateway-independent.

\begin{table}[t]
  \centering
  \caption{Prompt tokens by mode. Identical across all three LLM families.}
  \label{tab:prompt-tokens}
  \renewcommand{\arraystretch}{1.15}
  \footnotesize
  \begin{tabular}{lcccc}
    \toprule
    Selection & \textsc{Sel} & \textsc{Leg} & \textsc{Sig} & \textsc{Body} \\
    \midrule
    \texttt{placeOrder}            & 197 & 249 & 297 & 604 \\
    \texttt{recommendEngine}       & 158 & 199 & 214 & 386 \\
    \texttt{pricingSurge}          & 158 & 199 & 213 & 364 \\
    \texttt{baseRepoCollapse}      & 144 & 186 & 181 & 193 \\
    \texttt{reviewCollapse}        & 132 & 161 & 164 & 195 \\
    \texttt{userCollapse}          & 108 & 134 & 143 & 159 \\
    \midrule
    Mean (all 6)                   & 150 & 188 & 202 & 317 \\
    Mean (no-collapse)             & 171 & 216 & 241 & 451 \\
    Mean (collapse)                & 128 & 160 & 163 & 182 \\
    \bottomrule
  \end{tabular}
\end{table}

\textsc{Gacp-sig} trades a modest token premium (+7.5\% overall) for
graph-grounded inclusion decisions. On collapse fixtures the cost is
near-parity (+1.5\%), because Phase~2 replaces $k$ sibling bodies with
one parent contract plus one delegate. On the $k{=}4$ case,
\textsc{Gacp-sig} actually \emph{saves} 2.7\% (186$\to$181 tokens).
On no-collapse fixtures, \textsc{Gacp-sig} admits more callees than
\textsc{Legacy}'s cap: \texttt{placeOrder} emits 7 vs.\ 5, including
two structurally proximate helpers that the cap drops (+19.3\%).
Where the cap is not binding (\texttt{recommendationEngine},
\texttt{pricingEngineSurge}), the two modes emit identical callee sets.

\textsc{Gacp-body} roughly doubles the prompt on no-collapse fixtures
(+109.3\%) by including full method bodies. On collapse fixtures,
Phase~2 limits the increase to +13.7\%.

\Cref{tab:collapse} shows the reshape on the three collapse fixtures.
Phase~2 fires on all three and replaces $k$ near-duplicate siblings
with the common ancestor plus one concrete delegate; this is
algorithm-only and LLM-independent. Prompt-token counts are identical
across all three LLM families for every fixture$\times$mode
combination, confirming that \gacp's output is a pure function of the
graph and the selection.

\begin{table}[t]
  \centering
  \caption{Phase~2 reshape: \textsc{Legacy} emits $k$ siblings; \gacp{} emits parent + delegate.}
  \label{tab:collapse}
  \renewcommand{\arraystretch}{1.2}
  \footnotesize
  \begin{tabular}{p{0.18\linewidth} p{0.32\linewidth} p{0.36\linewidth}}
    \toprule
    Fixture & \textsc{Legacy} emits & \textsc{Gacp} emits \\
    \midrule
    \texttt{base\-Repo} ($k{=}4$)
      & 4 sibling repositories
      & \texttt{Base\-Repository} + \texttt{Customer\-Repository} \\
    \texttt{review\-able} ($k{=}3$)
      & \texttt{MenuItem}, \texttt{Restaurant}, \texttt{Driver}
      & \texttt{Reviewable} + \texttt{MenuItem} \\
    \texttt{abstract\-User} ($k{=}2$)
      & \texttt{Customer}, \texttt{Driver}
      & \texttt{Abstract\-User} + \texttt{Customer} \\
    \bottomrule
  \end{tabular}
\end{table}

%% file: sections/06-limitations.tex
\section{Limitations and Future Work}
\label{sec:limitations}

\paragraph{Single benchmark project}
The evaluation draws from one subject system, \textsc{quickbite}, chosen
for its seeded clone groups at three sizes ($k = 2, 3, 4$). The
algorithm is project-agnostic; running it on additional codebases is on
our follow-on agenda.

\paragraph{Coarse accuracy rubric}
The keyword-match accuracy score saturates to 2/2 across all modes and
LLM families. Quality claims in \Cref{sec:evaluation} therefore rest on
the algorithm-output comparison (callee-set composition and collapse
behaviour), which is LLM-independent. A behaviour-test rubric would
discriminate more finely but is follow-on work.

\paragraph{Token estimator}
Prompt tokens are estimated as character-count/4. Absolute counts under
provider-specific tokenizers differ by $\pm$10--15\%, but relative
comparisons, which are the basis of every claim, are preserved.

\paragraph{No human-comprehension claim}
This paper measures the algorithm's effect on the prompt, not whether
developers comprehend code faster. A controlled user study on
\textsc{quickbite} is designed but postdates this submission.

\paragraph{Personalization and graph granularity}
\gacp{} treats every developer as equivalent and operates on a
class-level dependency graph. Method-level call graphs and
expertise-adaptive pruning are future directions.

\paragraph{Baseline and parameter scope}
\Cref{sec:evaluation} compares \gacp{} against Selection-only and a
fixed-cap Legacy baseline, not against an embedding-retrieval system
(e.g., Cody, RepoFuse); a head-to-head comparison on the same six
fixtures is future work. We likewise did not sweep \gacp{}'s design
parameters ($K_{\max}$, edge-type weights); confirming that collapse
behaviour is stable under alternative settings is a natural next step.

%% file: sections/07-conclusion.tex
\section{Conclusion}
\label{sec:conclusion}

This paper presented \toolname, an Eclipse plugin that integrates
dependency-graph visualization, LLM-assisted code explanation, and
clone-refactoring discovery onto a single graph substrate. The 
unifying design principle is that the visualization the developer sees
is the data structure the tool reasons over: \gacp{} queries the graph
for callee distances, the clone overlay paints onto the graph, and the
developer reads the graph to verify both. \gacp's callee-inclusion 
decisions are auditable against visible graph nodes and reproducible
across LLM families, distinguishing \toolname{} from
embedding-retrieval and perplexity-based approaches whose relevance
signals are model-internal.

On the \textsc{quickbite} benchmark, \gacp's inheritance-collapse
phase reshapes prompts on clone-touching selections from $k$~sibling
duplicates to a parent-contract-plus-delegate form, and the
graph-distance cutoff admits structurally proximate helpers that a
fixed cap drops. The plugin, benchmark fixtures, and replay harness
are publicly available~\cite{comprendiaArtifact}.